\documentclass[aps,prb,reprint,groupedaddress,showpacs]{revtex4-1}

\usepackage[german, english]{babel}

\usepackage[utf8]{inputenc}
\usepackage[T1]{fontenc}

\usepackage{natbib}

\usepackage{amsmath}
\usepackage{amssymb}
\usepackage{bm}

\usepackage{textcomp}

\usepackage[pdftex]{graphicx}

\newcommand{\ie}{i.\,e.}
\newcommand{\eg}{e.\,g.}

\newcommand{\dd}{\ensuremath{\mathrm{d}}}

\newcommand{\rs}{\ensuremath{R_{\square} }}
\newcommand{\km}{\ensuremath{k_m}}
\newcommand{\ks}{\ensuremath{k_s}}
\newcommand{\kb}{\ensuremath{k_b}}
\newcommand{\Ts}{\ensuremath{T_s}}

\newcommand{\vecj}{{\ensuremath{\bm{\mathrm{j}}}}}
\newcommand{\vecr}{{\ensuremath{\bm{\mathrm{\bm{r}}}}}}
\newcommand{\vecn}{{\ensuremath{\bm{\mathrm{n}}}}}

\newcommand{\Tmax}{\ensuremath{T_{\text{max}}}}

\begin{document}

\title{Two-dimensional simulations of temperature and current-density
distribution in electromigrated structures}

\author{Birgit Kießig}
\affiliation{Karlsruhe Institute of Technology, Institut für
Festkörperphysik, D-76021 Karlsruhe, Germany}
\affiliation{ Karlsruhe Institut of Technology, Physikalisches Institut,
D-76131 Karlsruhe, Germany }

\author{Roland Schäfer}
\email{Roland.Schaefer@kit.edu}
\affiliation{Karlsruhe Institute of Technology, Institut für
Festkörperphysik, D-76021 Karlsruhe, Germany}

\author{Hilbert v.~Löhneysen}
\affiliation{Karlsruhe Institute of Technology, Institut für
Festkörperphysik, D-76021 Karlsruhe, Germany}
\affiliation{ Karlsruhe Institut of Technology, Physikalisches Institut,
D-76131 Karlsruhe, Germany }

\date{April 8, 2013}

\begin{abstract}
We report on the application of a feedback-controlled electromigration
technique for the formation of nanometer-sized gaps in mesoscopic gold
wires and rings.  The effect of current density and temperature, linked via
Joule heating, on the resulting gap size is investigated.  Experimentally,
a good thermal coupling to the substrate turned out to be crucial to reach
electrode spacings below 10\,nm and to avoid overall melting of the
nanostructures.  This finding is supported by numerical calculations of the
current-density and temperature profiles for structure layouts subjected to
electromigration.  The numerical method can be used for optimizing the
layout so as to predetermine the location where electromigation leads to
the formation of a gap.
\end{abstract}

\pacs{66.30.Qa, 73.63.RT, 44.10.+i, 02.70.Bf}

\maketitle

\section{\label{sec:intro}Introduction}
Electromigration leads to the formation of voids and hillocks in metallic
conductors by thermally assisted ion diffusion under the influence of an
applied electric field\cite{ho89}.  For more than a decade\cite{park99},
this process has been employed to deliberately form gaps of the order of a
few nanometers only in microstructures.  These gaps can be used to
incorporate molecules in a source-drain channel.  The planar geometry
facilitates the integration of gating electrodes that are essential for a
systematic characterization of electronic transport through molecules.

Initially, electromigration was performed at cryogenic temperatures and
reported to yield irregularly shaped slits with a narrowest separation $a$.
From low-temperature current-voltage measurements, $a\sim1$\,nm was
estimated\cite{park99, park00}.  However, metallic clusters often remained
in the generated gap\cite{houck05, heersche06, zant06}.  By working at room
temperature, the cleanliness of the gaps could be improved, while the
simultaneous implementation of a feedback-control still allowed the
generation of nanometer-spaced electrodes\cite{strachan05, esen05,
strachan07}.  We used a realization of this latter technique aimed at the
control of the power dissipated while a gap is being formed.  The procedure
is similar to the one described in Ref.\ \onlinecite{hoffmann08}.  Our aim
is to extend the technique used so far for gaps in single wires to more
complicated structures, \eg, to form gaps in the two arms of a ring
connected by leads at opposite sides.

To obtain a high yield of nanogaps, the experimental parameters have to be
optimized with respect to two competing objectives.  On the one hand, the
temperature $T$ and the current density \vecj{} have to be raised
sufficiently for material transport to occur on a reasonable time scale.
On the other hand, Joule heating has to be limited to avoid melting of the
metal structure, which would lead to droplet formation due to surface
tension and to an unwanted increase in electrode spacing.

In our experiments, we find a clear dependence of the resulting gap size on
the thermal coupling of the electromigrated structure to the cold
substrate.  We present a numerical model which leads to a plausible
explanation of this finding.  It maps the current density and the excess
temperature in the voltage-biased structure taking into account the heat
flow to the substrate.  The calculations can be used to optimize the sample
geometry so as to predetermine the location where electromigration leads to
the formation of voids and the structure is finally going to break. 

\section{\label{sec:exp}Experimental Observations}
All samples used for this work were prepared by standard e-beam lithography
on commercially available, slightly p-doped (boron, resistivity:
$1\cdots20\,$\textohm{}cm) silicon substrates.  Their surface was either
covered by 400\,nm thermally grown SiO$_2$ or merely by native silicon
oxide.  Here, we focus on samples made from 30\,nm thick gold deposited by
electron-beam evaporation without any additional adhesion layer.

Electromigration was performed at room temperature and mostly at ambient
pressure.  Some experiments were carried out in the high vacuum
($10^{-3}\,$Pa) of a scanning electron microscope (SEM) for in-situ
monitoring of the electromigration process.  The resulting gap size,
however, did not depend on the background pressure.  The active
feedback-control loop was realized by ramping a voltage across the sample
and measuring its resistance in a four-point configuration.  Upon detection
of a certain change of $\Delta R/R\sim 5\dots10$\,\%, the voltage was set
back to a safe value immediately by the control software and the voltage
ramp was reinitialized.  At an advanced stage of the process a sudden
change of resistance by several orders of magnitude indicates rupture of
the metallic current path and the procedure is stopped.

\begin{figure}
(a)\hspace{0.47\linewidth}(b)\hfill\phantom{.}\\[0mm]
\includegraphics[width=0.48\linewidth]{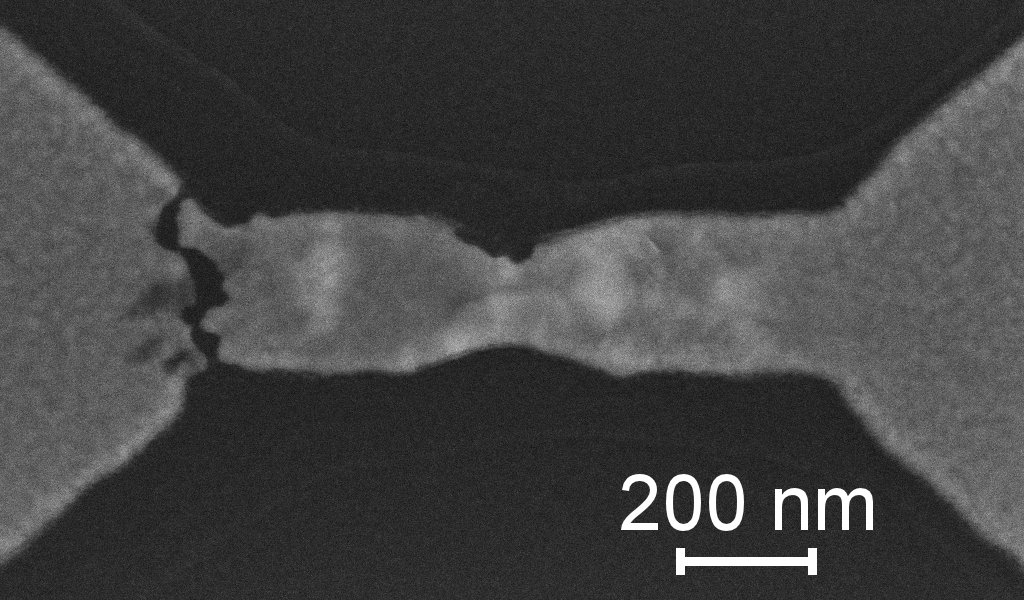}
\hfill
\includegraphics[width=0.48\linewidth]{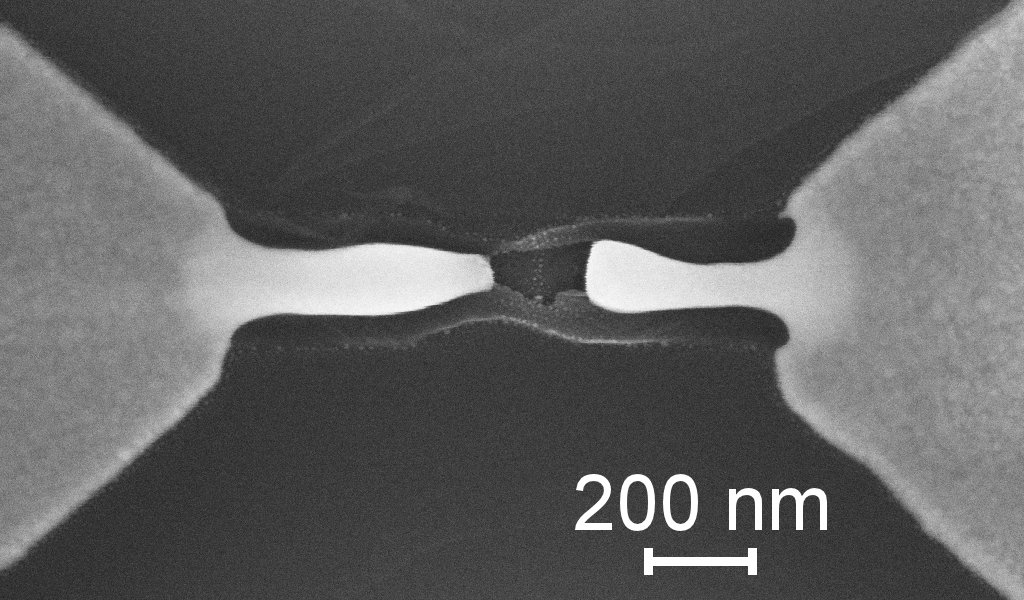}
\caption{\label{fig:i}SEM images of gold wires after electromigration in
high vacuum ($10^{-5}$\,mbar).  The substrate consists of p-doped silicon
covered by (a) native SiO$_x$ and (b) 400\,nm thermally grown SiO$_x$,
respectively.}
\end{figure}
In this paper we want to address the issue of Joule heating during
electromigration.  Its effect manifested itself mainly in two aspects of
our experiment.  On the one hand, SEM imaging after gap formation revealed
that metallic structures on silicon wafers covered by a 400\,nm thick
thermally grown silicon oxide layer tend to show signs of melting.
Consequently, the resulting electrode separation is considerably above
10\,nm.  A typical example is shown in Fig.\ \ref{fig:i} (b).  Samples
electromigrated on silicon covered by native oxide, however, display the
intended formation of voids and hillocks leading to the opening of an
irregular gap, whose expansion lies below the resolution of the SEM, \ie,
$a\lesssim 10$\,nm.  An example is shown in Fig.\ \ref{fig:i} (a).  On the
other hand, as can be seen in the same figure, the samples were usually
disrupted near their cathode end and not at the constriction prestructured
for this purpose at the center of the wire.

Breaking at the cathode side is a common observation in \emph{straight}
wires\cite{blech67, blech69, ho89, heersche07}.  It arises because
electromigration is due to two mechanisms leading to atom transport in
opposite directions.  First, atom transport it is directly driven by the
electrostatic force acting on the positively charged atom cores in the
applied electric field.  This mass transport is directed towards the
cathode.  Second, the cores are driven by the so called ``\textit{electron
wind}'', \ie, the momentum transfer from the conduction electrons to
defects in inelastic scattering events.  The net contribution of these
impacts is parallel to the electron current.  In gold the motion due to the
electron wind predominates and thus material is effectively transported
towards the anode.  Since the amount of material flow depends strongly on
temperature, electromigration is enhanced by Joule heating in those parts
of the sample where the current density is highest and leads to depletion
of material at the cathode side of narrow constrictions in the current
path. 

The purpose of the constriction shown in the center of Fig.\ \ref{fig:i}
(b) is to enhance the current density and create a hot spot at a predefined
position.  Electromigation was expected to lead to a break junction at this
very location which, however, was not observed.  Apparently, the conditions
at the lead-nanowire-junction have to be taken into account more carefully.
In the next section we present a model to calculate the temperature profile
for a given sample layout which takes into account the heat flow from the
metal film to the leads as well as to the substrate.  The latter part is of
utmost importance as indicated by our experimental observations presented
in Fig.\ \ref{fig:i}.

\section{\label{sec:calc}Calculation of the heat distribution}
The temperature in a voltage-biased sample is enhanced due to Joule
heating.  We present a model to quantify this effect.  It is applicable to
metallic structures of constant thickness $d$, sheet resistance \rs, and
thermal conductivity \km.  Heat is carried along the metal film to the
leads and through an insulating barrier of thickness $b$ and thermal
conductivity \kb, where the condition $\kb\ll\km$ naturally holds.  The
structure is supported below the barrier by a thick substrate with thermal
conductivity \ks.  The substrate is assumed to be thermally well coupled at
the back side to a heat reservoir at ambient temperature.  It is heated
from the top by the power flowing through the barrier which results from
the dissipation in the metallic structure.  The situation in the substrate
is therefore intrinsically three dimensional.  However, we can safely
assume that the temperature in the metallic structure itself does not vary
significantly perpendicular to the film plain (owing to $\kb\ll\km$).  The
temperature rise in the substrate depends strongly on the barrier thickness
$b$ and is significant for small $b$ only.  We postpone the discussion of
thicker barriers to the end of this section and assume first that the
temperature varies linearly across the insulating barrier, \ie, we consider
a laminar heat flow across the insulator perpendicular to its interfaces
with the metal film and the substrate.  The numerics for the metallic film
is reduced to a two-dimensional problem obeying the modified Poisson
equation:
\begin{equation}\label{equ:poisson}
\km{}d\nabla^2T+\dot{q}=\frac{\kb}{b}(T-\Ts).
\end{equation}
Here $T(\vecr)$ is the temperature elevation relative to the base
temperature of the unbiased sample, $\dot{q}=\mathrm{\mathbf{j}}^2\rs$ the
power density, $\vecj(\vecr)$ the two-dimensional current density, and
$\Ts(\vecr)$ the temperature at the interface between of the substrate and
the barrier.  The left-hand side of Eq.\ (\ref{equ:poisson}) describes the
heat balance in the metallic film, while the right-hand side is equal to
the loss due to heat flow to the substrate.  The latter depends implicitly
on $\Ts(\vecr)$ which itself obeys the three-dimensional Laplace equation
(describing the source-free heat flow in the thick substrate)
\begin{equation}\label{equ:laplace}
\nabla^2\Ts=0
\end{equation}
subjected to mixed boundary conditions.  At the interface with the
insulating barrier we have
\begin{equation}\label{equ:vneumann}
\nabla_\vecn{}\Ts=\frac{\kb}{b\ks}(\Ts-T),
\end{equation}
where $\nabla_\vecn$ is the gradient perpendicular to the interface.  At
all other boundaries we approximate $\Ts(\vecr)$ by zero.  Equations
(\ref{equ:poisson}) and (\ref{equ:laplace}) form a set of mutually
dependent relations. 

\begin{figure}
\includegraphics[width=\linewidth]{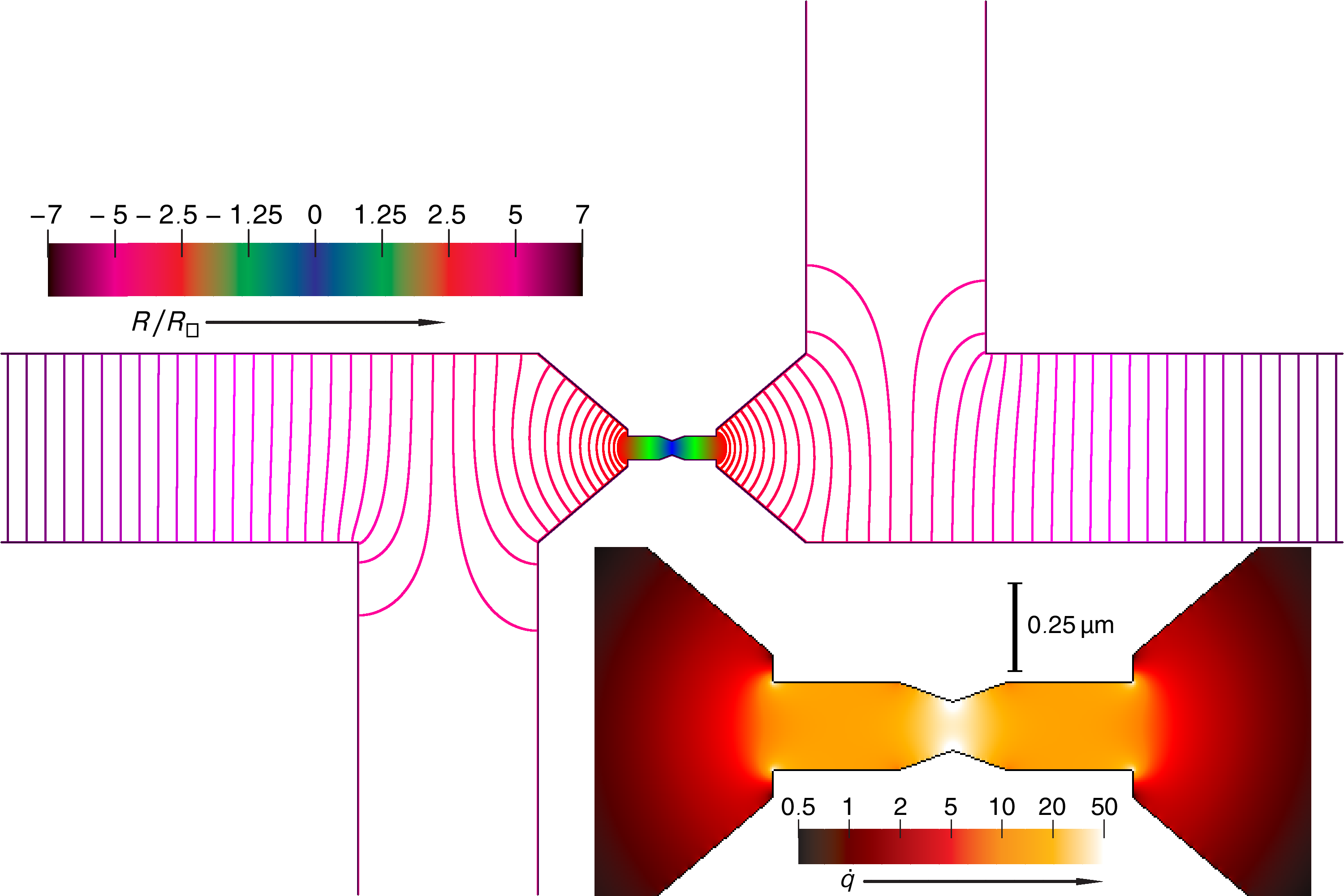}
\caption{\label{fig:ii}Equipotentential lines for a test layout biased in
horizontal direction.  The taps pointing up- and downwards represent
voltage probes.  The voltage drop between consecutive lines is
$\Delta{}V=0.1\rs\cdot{}I$ and the total drop between the outermost lines
$V=12.4\rs\cdot{}I.$ The inner region where the potential lines merge to a
continuous color scale corresponds to $\Delta{}V=5\rs\cdot{}I.$ The inset
displays the resulting power density $\dot{q}$ in units of
$I^2\rs/$\textmu{}m$^2$ for the center of the main figure.  Notice the
peaking of the power density at concave corners.  }
\end{figure}

We find it convenient to solve this set of equations by a finite-difference
method on a square lattice and discretize the relevant parts of the
metallic structure with a predefined cell size sufficiently small to
capture all fine details.  For this purpose we use an appropriate
rectangular cutout of the e-beam lithography design file, which covers the
area of highest current densities.  For the examples presented below, a
cell size between 1\,nm and 5\,nm turned out to be reasonable, with several
million cells within the cutout.  The current and voltage leads (cf.\
horizontal and vertical structures, respectively, emanating from the center
in Fig.\ \ref{fig:ii}) have constant cross section and extend beyond the
cutout.  In addition the length of the leads included in the explicit
calculation exceeds their width.  These measures guarentee that the
equipotential lines beyond the cutout generated by an electromotive force
between source and drain current leads run perpendicular to the latter.

Before solving Eqs.\ (\ref{equ:poisson}) and (\ref{equ:laplace}) we have to
calculate the current density \vecj\ which enters the second term on the
left-hand side of Eq.\ (\ref{equ:poisson}).  \vecj\ is directly related by
$\vecj=\nabla{}V/\rs$ to the potential $V$, which in turn obeys the Laplace
equation $\nabla^2V=0$ with mixed boundary conditions.  We solve the latter
equation by successive overrelaxation (SOR)\cite{press07}.  For edges of
the metallic structure internal to the explicitly treated area as well as
for sections of voltage probes with its border we have von-Neumann
conditions, $\nabla_\vecn{}V=0$.  At the section of the source and drain
leads with the border we apply the actual bias $V_\text{max}$ and
$V_\text{min}$, respectively.  After solving the Laplace equation subjected
to these boundary conditions, the total current
$I=\int\dd{}s|\nabla_\vecn{}V|/\rs$ can be calculated, where the integral
runs along an arbitrary line cutting the source-drain path.  It is
convenient to normalize $V$ by $I\rs$, since $V/I\rs$ directly reflects the
``\emph{number of squares}'' between different points within the metallic
structure, \ie, the length-to-width ratio of a strip which would acquire an
equivalent voltage drop if biased by the same current.  Figure \ref{fig:ii}
shows the result of the calculation for a layout typically used in our
experiments.  The inset of this figure displays the power density in the
center of the structure.  The notch in the middle does indeed lead to the
intended peak of the power density at this position (and of the current
density that drives electromigration).  However, a similar peak is found at
the crossover from the funnel-shaped leads to the narrow wire as indicated
by the four lightly colored spots at the rectangular corners in the inset
of Fig.  \ref{fig:ii}.  This is due to the well-known divergence of the
electric field strength at corners\cite{jackson99}.  In our numerics, this
divergence is limited only by the finite cell size used in the
finite-difference method as corners are infinitely sharp in the layout.  In
practice, e-beam lithography leads to a minimal radius at corners set by
the finite resolution of the process and the granularity of the crystalline
metallic layers.  This radius sets the scale of the actual current-density
enhancements.  It can thus be expected that the current density reaches the
same order of magnitude at each concave corner.  The precise value depends,
however, on the opening angles and can be tuned by proper design.
Consequently, rectangular corners as the ones visible in Fig.\ \ref{fig:ii}
should be avoided in the strongly heated parts of the structure, unless a
starting point for electromigration is intended.  We are going to comment
further on this issue in the next section.

Having $\dot{q}=(\nabla{}V)^2/\rs$ at hand, we solve Eqs.\
(\ref{equ:poisson}) and (\ref{equ:laplace}) simultaneously by an iterative
procedure.  Starting with $T(\vecr)=\Ts(\vecr)=0$ we first relax Eq.\
(\ref{equ:poisson}) at constant {\Ts} to get an updated estimate for
$T(\vecr)$.  The boundary conditions for the structure edges are again of
von-Neumann type, $\nabla_\vecn{}T=0$.  Special care has to be taken to
treat the heat flow at the border of the cutout.  Owing to the
aforementioned homogeneous nature of the leads extending beyond this
border, the heat-flow problem is essentially one dimensional
here\cite{durkan07}.  In long leads the temperature along the wire
direction approaches exponentially a steady-state value of
\[T_a=\Ts+I_\ell^2\rs{}b/(w^2\kb),\] where $I_\ell$ is the current in the
specific lead of width $w$.  In general, the substrate temperature {\Ts} in
this equation has to be determined self-consistently as it results from the
heat flow caused by the last term.  However, {\Ts} is naturally small
compared to more strongly heated zones below spots of high current density.
We ignore small variations of {\Ts} at the border of the cutout and take in
the above formula for {\Ts} a self-consistently calculated value.  The
functional dependence of the temperature along the leads is then given by 
\begin{equation}\label{equ:onedt}
T(x)=T_a+(T_0-T_a)\exp(-(x-x_0)/\lambda).
\end{equation}
Here $x$ is a spatial coordinate along the current direction entering the
cutout at $x_0$, $T_0$ is the temperature of the cross section at $x_0$,
and $\lambda=\sqrt{bt\km/\kb}$ is the length scale of relaxation towards
the asymptotic $T_a$.  From Eq.\ (\ref{equ:onedt}) the boundary condition
of Eq.\ (\ref{equ:poisson}) at the cutout frame can be deduced, 
\begin{equation}\label{equ:bvpoisson}
\nabla_\vecn{}T=(T_a-T_0)/\lambda.
\end{equation}

Eq.\ (\ref{equ:bvpoisson}) completes the boundary-value problem defined in
Eq.\ (\ref{equ:poisson}) and is used in a successive overrelaxation loop to
update $T(\vecr)$.  In the next step of the iterative procedure we fix
$T(\vecr)$ in Eq.\ (\ref{equ:vneumann}) and update $\Ts(\vecr)$ via Eq.\
(\ref{equ:laplace}).  Here, successive overrelaxtion would fail since the
3d nature of heat relaxation in the substrate leads to an intractably large
number of cells in the discretization step.  The final goal of our
calculation, however, is the determination of $T(\vecr)$ defined in the 2d
plane of the metallic structure.  To this end we need to know $\Ts(\vecr)$
at the interface between the substrate and the barrier only.  Let
$G(\vecr)$ be the solution of $\nabla^2G=0$ in the 3d half-space of $z<0$
obeying Dirichlet conditions $G(\vecr\to\infty)=0$ and von-Neumann
conditions at $z=0$, $\nabla_\vecn{}G=\delta(\vecr)$.  Due to the linearity
of the Laplace equation we can write down immediately the solution of Eq.\
(\ref{equ:laplace}) obeying the boundary condition of
Eq.~(\ref{equ:vneumann}),
\begin{equation}\label{equ:solulp}
T_s(\vecr)=\frac{\kb}{b\ks}\int\dd{}\vecr^\prime{}G(\vecr-\vecr^\prime)
\left(T(\vecr^\prime)-\Ts(\vecr^\prime)\right).
\end{equation} 
The integral is restricted to the plane $z=0$.  Thus, the calculation of
$\Ts(\vecr)$ at the interface $z=0$ requires the knowledge of $G(\vecr)$ at
the interface only as well.  In this way Eq.\ (\ref{equ:solulp}) is reduced
to a two-dimensional convolution which can be solved numerically by
discrete Fourier transforms (DFT) for which fast algorithms exist.  Our
problem is thus reduced to the determination of $G(\vecr)$.  Details of
this calculation are given in the appendix.

Eq.\ (\ref{equ:solulp}) can now be used to update $\Ts(\vecr)$.  Note,
however, that Eq.\ (\ref{equ:solulp}) does define {\Ts} only implicitly.
Therefore a further relaxation loop is required.  In each iteration of this
loop the convolution on the right-hand side is calculated employing the
convolution theorem by two consecutive DFT steps.  The result is
subsequently used to update \Ts.  In the course of the iteration, the
changes in {\Ts} get smaller and the loop is terminated after sufficient
accuracy has been obtained.  This ends the first round of our overall
iterative procedure.  In the following iterations we proceed by using Eq.\
(\ref{equ:poisson}) at constant {\Ts} to update $T$ and Eq.\
(\ref{equ:laplace}) at constant $T$ to update {\Ts}.  The procedure
converges to a self-consistent solution after several iterations. 

\begin{figure*}[t]
\includegraphics[width=\textwidth]{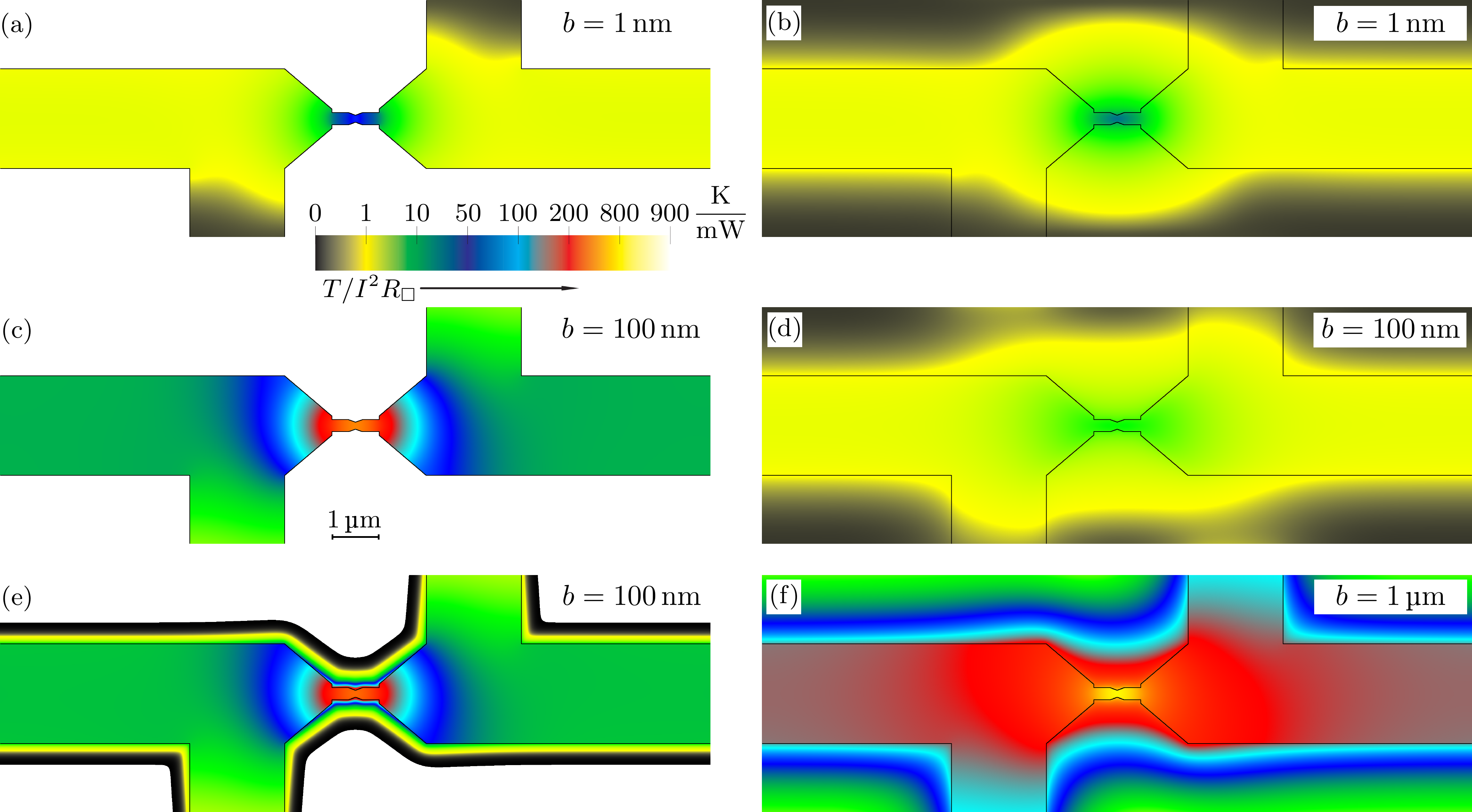}
\caption{\label{fig:iii} Maps of normalized temperature, $T/I^2\rs,$ of
metallic structure [(a) and (c)], substrate-barrier interface [(b) and
(d)], and interface between barrier and metallic structure [(e) and (f)].
Colors are according to the scale in units of K/mW.  Results for three
barrier thicknesses are shown: $b=1\,$nm [(a) and (b)], $b=100\,$nm [(c),
(d), and (e)], and $b=1\,$\textmu{}m (f). (a) to (d) are calculated with
Approx.\ I, while for (e) and (f) Approx.\ II is used.}
\end{figure*}
So far we have developed a model applicable for thin barrier thickness $b$.
From now on, we refer to this particular approximation as Approx.\ I.  In
the next section we present results showing that indeed substrate
temperature {\Ts} and $T$ follow each other closely for small $b$ and the
\emph{a-priori} approximation of a laminar heat flow perpendicular to the
interface between substrate and barrier is justified in this case.
However, {\Ts} is a monotonically decreasing function of $b$ while at the
same time $T$ rises monotonically Moreover, at small $b$ it is possible to
heat the substrate rather locally by the large power density resulting from
a notch as, \eg, the one in the center of the test structure displayed in
Fig.\ \ref{fig:ii}.  If $b$ is increased the power is delivered to the
substrate more evenly and the temperature distribution tends to smooth out
(with a small enhancement only even at the center).  The temperature
difference $T-\Ts$ soon becomes so large that {\Ts} on the right-hand side
of Eq.\ (\ref{equ:poisson}) can safely be neglected.  In this case, the
vertical nature of the heat flow is lost and the problem of the heat
distribution in the isolating barrier has to be treated more precisely.
Fortunately, this can be done by the same method introduced so far.  For
thicker barriers we take the temperature at the interface of the substrate
with the barrier as unelevated, the barrier now stays cold at its back side
and takes thus the role formerly played by the substrate.  The set of Eqs.
(\ref{equ:poisson}), (\ref{equ:laplace}), and (\ref{equ:vneumann}) now
reads
\begin{eqnarray}\label{equ:poissonlaplace}
\nonumber\km{}d\nabla^2T+\dot{q}=\kb\frac{\Delta{}T}{\varepsilon},\quad
\nabla^2\Ts&=&0\\*
\nabla_\vecn\Ts&=&-\frac{\Delta{}T}{\varepsilon},
\end{eqnarray}
where $\Delta{}T$ is the temperature drop across a slice of the barrier of
infinitesimal thickness $\varepsilon$.  In the finite-difference treatment
of the problem we set $\varepsilon$ equal to the discretization cell size.
In Eq. (\ref{equ:poissonlaplace}) $\Ts=\Ts(\vecr)$ represents the 3d
temperature profile in the \emph{barrier}.  As such, it is defined at the
interface of the barrier with the metal film, too.  At this interface,
\Ts{} equals $T$ (in the numerics it differs marginally since the
infinitesimal $\varepsilon$ is approximated by a variable of finite
extent).  The thickness $b$ enters this approximation via a modification of
$G(\vecr)$  in Eq.\ (\ref{equ:solulp}), which we label $G_b(\vecr)$.  It
still obeys the Laplace equation $\nabla^2G_b=0$, now in the barrier, \ie,
for $-b<z<0$, but the Dirichlet condition $G(\vecr\to\infty)=0$ is replaced
at $z=-b$ by $G_b(z=-b)=0$.  In this way we approximate the temperature at
the interface between barrier and substrate by the base temperature of the
unbiased sample.  We refer to this type of approximation as Approx.\ II in
the following section.  The calculation of $G_b(\vecr)$ on a rectangular
grid used in the finite-difference approximation is explained in the
appendix.

\section{\label{sec:res} Results}
The model developed in the last section can be used to optimize structure
layouts intended for the formation of narrow break junctions at well
defined locations by deliberate electromigation.  In this section we
present some examples.

First we want to address the question why metallic structures on a thicker
silicon oxide layer are more likely to melt during the electromigration
process than samples on native SiO$_2$.  For that purpose we calculate the
excess-temperature map for the sample geometry shown in Fig.\ \ref{fig:ii}
for varying barrier height $b$.  For definiteness we consider gold
structures ($\km\approx316\,$W/Km\cite{powell66}) with a layer thickness of
$d=30\,$nm, approximate the thermal conductivity of the substrate by the
typical value of silicon at room temperature,
$\ks\approx150\,$W/Km\cite{glassbrenner64}, and vary $b$ between 1\,nm and
2.5\,\textmu{}m.  For the thermal conductivity of the barrier we use that
of thin SiO$_2$ films at room temperature,
$\kb\approx1.1\,$W/Km\cite{kleiner96}. 
\begin{figure}
\includegraphics[width=\linewidth]{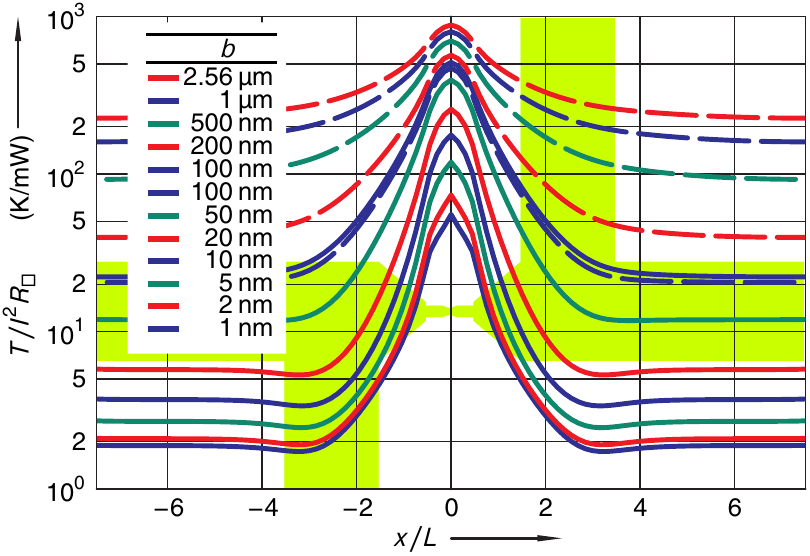}%
\caption{\label{fig:iv}Excess temperature $T$ along the central axis of the
same structure as used in Fig.\ \ref{fig:iii}.  Note the logarithmic
vertical scale.  A picture of the structure has been underlayed to ease
matching regions of different current densities with the shown curves.
$L=1$\,\textmu{}m is defined as the length of the narrow wire in the
center.  The thickness $b$ of the SiO$_2$ barrier is varied between 1\,nm
and 2.56\,\textmu{}m, the temperature is normalized by $I^2\rs$.  Solid
lines correspond to Approx.\ I, while dashed lines are calculated using
Approx.\ II.}
 \end{figure}
\begin{figure}
(a)\hspace{0.47\linewidth}(b)\hfill\phantom{.}\\[0mm]
\includegraphics[height=0.33\linewidth]{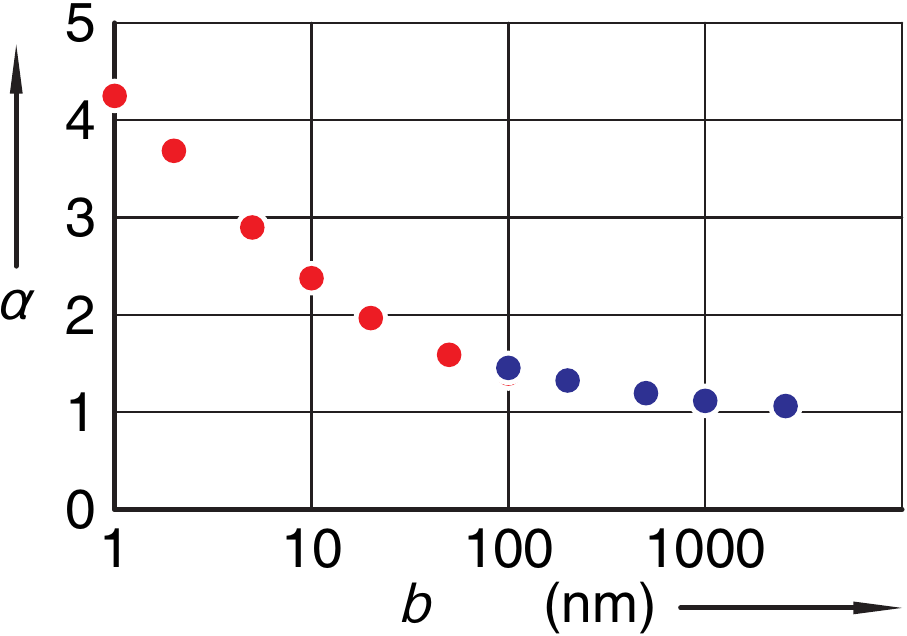}\hfill%
\includegraphics[height=0.33\linewidth]{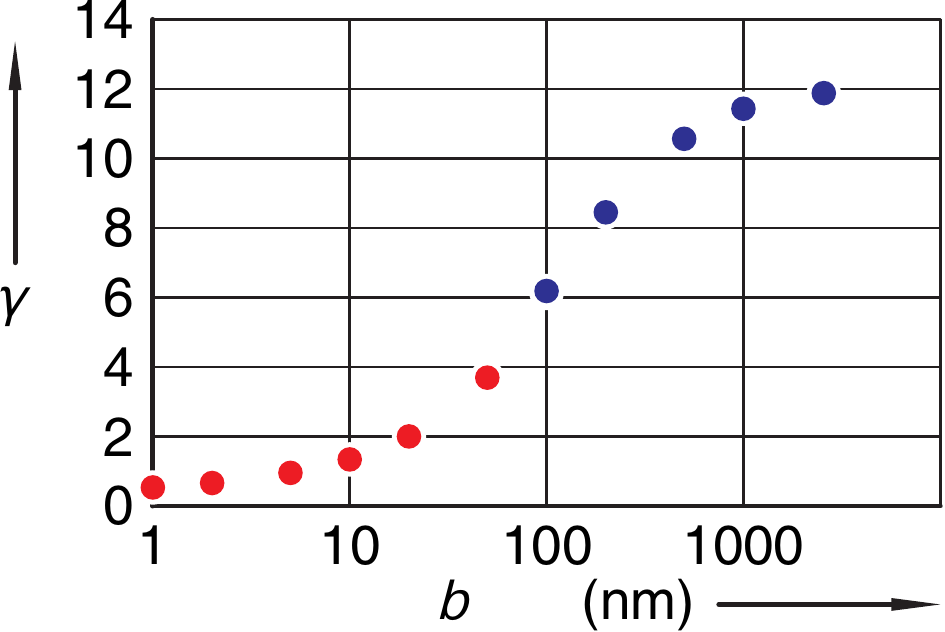}\\
(c)\hspace{0.47\linewidth}(d)\hfill\phantom{.}\\[0mm]
\includegraphics[height=0.33\linewidth]{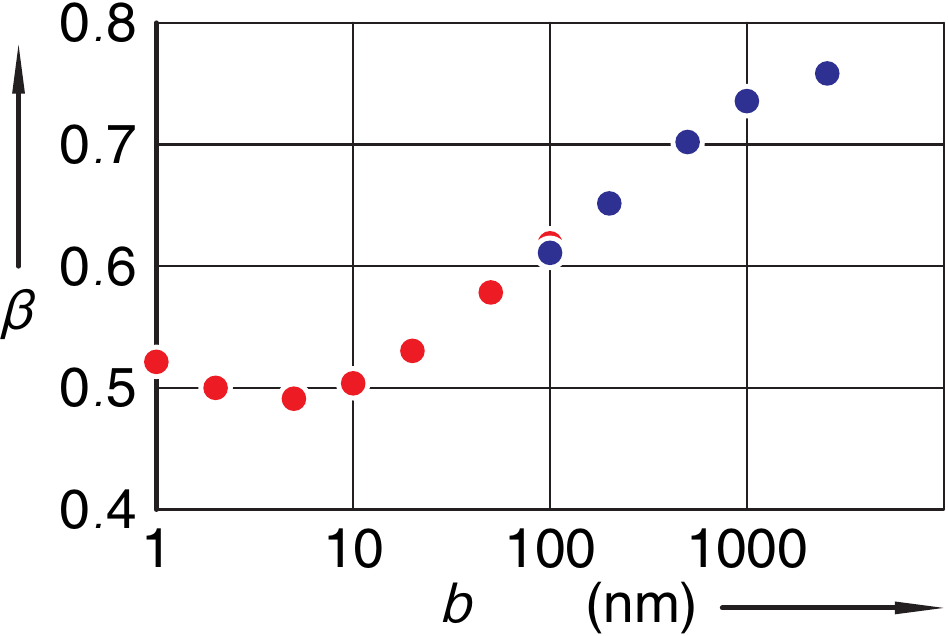}\hfill%
\includegraphics[height=0.33\linewidth]{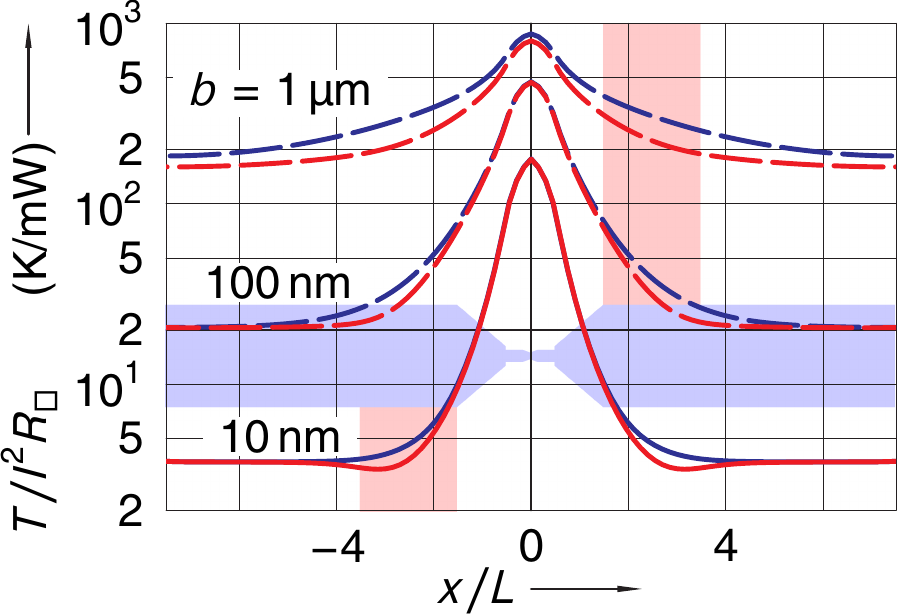}
\caption{\label{fig:v}(a) Plot of $\alpha=I\sqrt{\rs/T_0}$ in units of
mA$\sqrt{\text{\textohm/K}}$ as a function of the barrier thickness $b$.
Here $I$ is the current needed to heat the structure of Fig.~\ref{fig:iii}
in its center to a temperature $T_0$.
(b) Ratio $\gamma$ of the first term in Eq.\ (\ref{equ:poisson}) to the
negative of the term on the right hand side of the same equation in the
center of the structure, $\gamma=(db\km/\kb)\nabla^2T/(\Ts-T)$.  This
compares the amount of heat flow in the metal to the effect of cooling by
the substrate.  At large $b$ the structure is thermally decoupled in its
center.
(c) Ratio $\beta=T_1/\Tmax$ of the temperature $T_1$ at the crossover from
the narrow wire to the funnel-shaped parts of the leads and the temperature
$\Tmax$ in the center as a function of $b$.
Red dots in (a), (b), and (c) are calculated with Approx.\ I, blue dots
with Approx.\ II.
(d) Temperature profile along the central axis of two different structures
at $b=10$, 100 and 1000\,nm.  The bluish and reddish underlay indicate
current and voltage probes, respectively.  Red lines correspond to results
for the four-point configuration shown in Fig.\ \ref{fig:iii} (complete
underlay).  Blue lines are calculated for a two point configuration, \ie,
the voltage probes are missing (bluish underlay only).  For $b=10\,$nm
Approx.\ I and for larger $b$ Approx.\ II has been used.}
\end{figure}

Figure \ref{fig:iii} shows the results for $b=1$\,nm, $b=100\,$nm, and
$b=1\,$\textmu{}m.  Obviously, the temperatures of substrate and metallic
structure follow each other closely for very thin barriers [see Fig.\
\ref{fig:iii} (a) and (b)].  For thicker barriers the asymptotic
temperature $T_a$ in the leads can easily reach values that exceed the
maximum temperature for thin barriers [compare Fig.\ \ref{fig:iii} (a) and
(c), calculated with Approx.\ I, \ie, a 3d heat relaxation in the substrate
and a vertical laminar flow in the barrier].  The temperature is maximal at
the center of the wire as expected.  A typical value for the sheet
resistance is $\rs=2\,$\textohm, so that $I^2\rs=1\,$mW at around
$I=20\,$mA.  For this current a moderate temperature rise of
$\Tmax\approx50\,$K results at the center of the narrow wire, if the
barrier thickness $b=1\,$nm.  For $b=100\,$nm the temperature rises already
by $\Tmax\approx510\,$K.  The substrate temperature below the leads,
however, does not change significantly for thicker barriers due to the fact
that the substrate underneath long leads is heated by the power density
$I^2\rs/w^2$ independent of barrier thickness [compare Fig.\ \ref{fig:iii}
(b) and (d)].  For $\rs=2\,$\textohm{} we get a marginal warming by about
1.6\,K at $I=20\,$mA.  Furthermore, substrate heating below the center of
the wire is reduced with increasing $b$.  Between $b=1\,$nm and $b=100\,$nm
the maximum temperature rise in the substrate drops from about 30\,K to
10\,K.  This is a rather low temperature increase as compared to the rise
in the metallic structure, \ie, Approx.\ II is a better choice for
$b\gtrsim100\,$nm.  Figure \ref{fig:iii} (e) displays the temperature
profile in the plain defined by the interface between the barrier and metal
film calculated in the framework of Approx.\ II, \ie\ evaluating the 3d
heat relaxation in the barrier, while the substrate is assumed to remain
cold.  Approximations I and II lead to similar results at $b=100\,$nm as
can be judged by inspecting Fig.\ \ref{fig:iii} (c) and (e).  Barrier
heating is restricted mainly to the regions underneath the metallic
structure for $b=100\,$nm with a small halo only.  At larger $b$ [see Fig.\
\ref{fig:iii} (f)], however, the area of the barrier with a significant
temperature rise has a considerably larger extent.  Nevertheless, a
pronounced maximum of the temperature in the central spot of the structure
remains the dominant feature.\footnote{Our model does not take into account
the effect of radiation cooling because the latter is negligible.  The
radiation losses are bounded by
$\sigma{}T_m^4\approx0.2\,$\textmu{}W/\textmu{}m$^2$ ($\sigma$ is the
Stefan-Boltzmann constant and $T_m=1337.33\,$K the melting point of gold)
and thus several orders of magnitude smaller than the power carried by heat
transport in case of significant heating.} 

Data for further values of $b$ are summarized in Fig.\ \ref{fig:iv}, where
the temperature profile along the central axis is displayed.  The
successive rise of the asymptotic lead temperature $T_a$ with $b$ for
$|x/L|\gtrsim4$ is the most prominent feature in this figure.  Furthermore,
one can observe that the ratio between the maximum temperature in the
center and $T_a$ is strongly reduced with increasing $b$.  For $b=100\,$nm
results of Approx.\ I and II are included.  Both approximations yield
similar results at this thickness, with slightly lower values of $T$ for
Approx.\ II due to the more efficient cooling resulting from the
temperature halo around the metallic structure visible in Fig.\
\ref{fig:iii} (e). 

In a well accepted approximation\cite{ho89}, the amount of gold pushed in
the anode direction is proportional to both the current density and the
self-diffusivity of the metallic cores.  The latter has a strong
temperature dependence described over a wide range by an Arrhenius
law\cite{lbnsiii26} (for self-diffusion data on Au see, \eg, Refs.\
\onlinecite{gupta73a,gupta73b,gupta74,mclean68}).  No matter whether
electromgration is dominated by diffusion in the bulk, along grain
boundaries, or at the surface, a sizeable effect sufficient for the
deliberate formation of voids requires elevated temperatures.  As a rule of
thumb, $T$ should be at least half the melting temperature $T_m$, in which
case electromigration along grain boundaries is the dominant atomic flux
mechanism.  The effect of electromigration is much stronger just below
melting and is then in general dominated by flux within the bulk.  However,
for a controlled process of gap formation the melting point must not be
exceeded.  In Fig.\ \ref{fig:v} (a) we plot $\alpha=\sqrt{1/y_\text{max}}$
as a function of $b$, where $y_\text{max}$ are the maximal values of the
curves in Fig.\ \ref{fig:iv} at $x=0$, where the temperature in the
metallic structure is highest.  If we fix the temperature at a value $T_0$
well below the melting point (let's say at $T_0\approx1000\,$K for gold)
$I_0=\alpha\sqrt{T_0/\rs}$ directly gives a safe current limit for the
specific structure.  This safe current limit is considerably reduced when
$b$ is increased as shown in \ref{fig:v} (a).  For a native oxide barrier
of $b\sim1\,$nm and $\rs\approx2\,$\textohm{} a current of almost $100\,$mA
can be tolerated.  For the thermally grown oxide layer of our experiments
($b=400\,$nm) the safe current limit is by a factor of four smaller.  This
makes controlled electromigration considerably more difficult.  The
feedback loop as implemented by us and others\cite{hoffmann08} uses a
resistance change to judge the appropriate current density.  This procedure
however might lead to current densities that exceed the limit set by the
melting point of gold. 

It is instructive to compare the size of the first term in Eq.
(\ref{equ:poisson}), \ie, the heat transport in the metallic structure, to
the size of the right-hand side of the same equation, \ie, the heat flow to
the substrate.  This is done in Fig.\ \ref{fig:v} (b) for the central point
$x=0$ of the structure where the temperature for a given current $I$ is
highest.  The ratio \[\gamma=(db\km/\kb)(\nabla^2T/(\Ts-T))\] of these two
terms is considerably smaller than $1$ for $b\lesssim5\,$nm, \ie, the
center of the structure is efficiently cooled via the substrate.  For
larger $b$, however, $\gamma$ rises quickly indicating that the coupling to
the substrate becomes increasingly inefficient and cooling has to rely on
the heat flow along the metallic leads.  The barrier thickness at which
this decoupling sets in is expected to be reduced with decreasing width of
the notch in the center.  Thus, if void formation by electromigration
occurs and the current path through a narrowing constriction, temperature
is dominated more and more by the heat flow along the structure.  This
leads to an even stronger tendency towards melting which has to be
counteracted by increasingly careful settings in the feedback-control loop
and necessitates good thermal coupling from the beginning.  In summary, our
analysis gives clear evidence why controlled electromigation on thick oxide
barriers is more likely to fail. 

Figure \ref{fig:v} (c) displays the ratio $\beta$ of the temperatures at
the wire end (\ie, at the crossover from the narrow region to the
funnel-shaped parts of the leads) and at the center, as a function of $b$.
Ideally, the temperature peaks so strongly at the center that
electromigration acts only there with reasonable strength.  This is
difficult to achieve, though.  Figure \ref{fig:v} (b) indicates that an
optimal barrier thickness $b_\text{opt}$ exists at which
$\beta(b_\text{opt})\approx0.5$.  This factor-of-two reduction of the
temperature is not sufficient as demonstrated by the experimental result
presented in Fig.\ \ref{fig:i} (a).  As indicated in the inset of Fig.\
\ref{fig:ii} the current density peaks not only at the notch in the center
but at concave corners as well.  Electromigration leads to material flux in
regions of high current density.  The formation of voids or hillocks, in
turn, requires some kind of inhomogeneity that causes a divergence of
material flux.  Such an inhomogeneity might be simply due to the
statistical distribution of grains comprising the metallic structure, but
regions of high temperature gradient lead for sure to a divergence in
atomic flux as the delivery of material from cold regions is less efficient
than delivery of material to the hotter parts.  In the structure studied
here the temperature gradient is large at the wire end and vanishes in the
center where the notch is located.  In the latter position material
transport tends to be balanced and thus the wire cross-section does not
change to first order.  Only at the corner at the cathode end void
formation is favored.

\begin{figure}
(a)\hfill\phantom{.}\\[0mm]
\includegraphics[width=\linewidth]{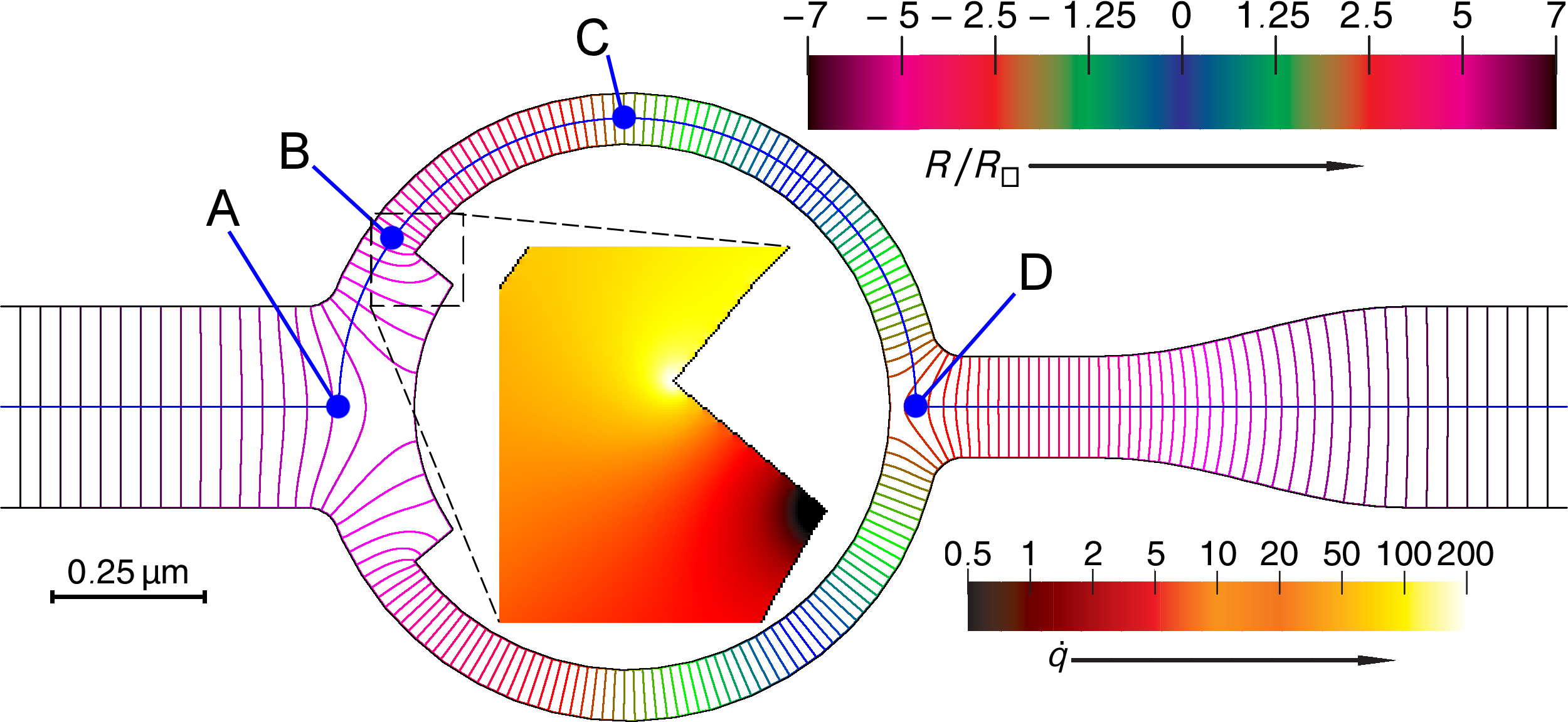}
(b)\hspace{0.47\linewidth}(c)\hfill\phantom{.}\\[0mm]
\parbox[t]{0.48\linewidth}{
\rule{\linewidth}{0pt}\\[-2mm]
\includegraphics[width=\linewidth]{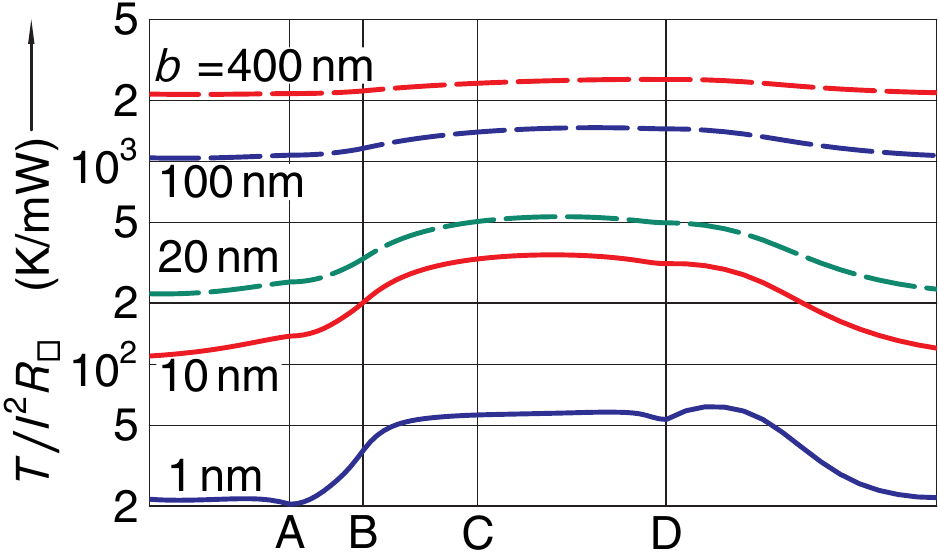}}
\parbox[t]{0.48\linewidth}{
\rule{\linewidth}{0pt}\\[-2mm]
\includegraphics[width=\linewidth]{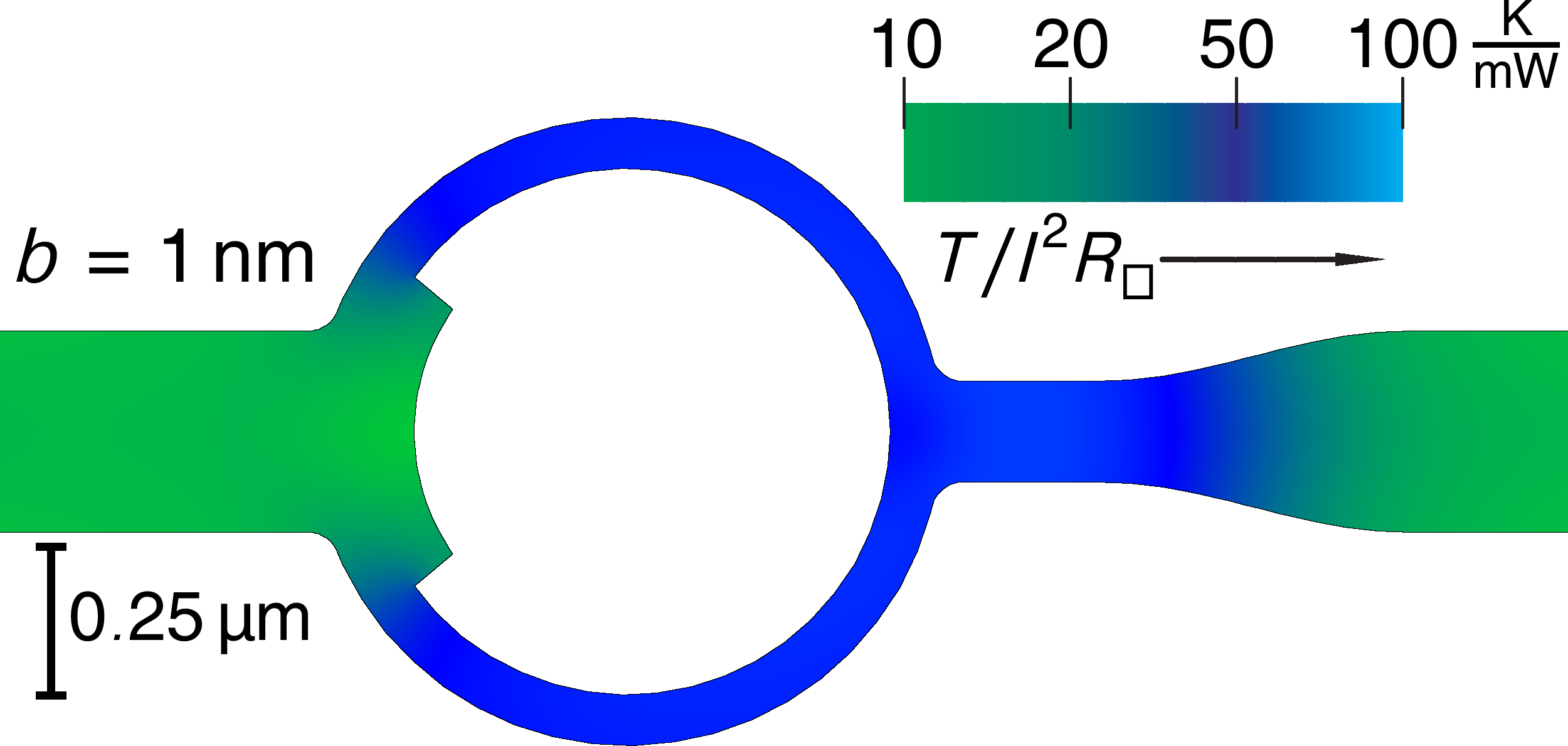}}
\caption{\label{fig:vi}
(a) Equipotential lines for a ring structure.  The voltage drop between
consecutive lines is $\Delta{}V=0.1R_\square\cdot{}I$ and the total voltage
drop between the outermost lines is $\Delta{}V=13.8R_\square\cdot{}I$.  The
enlarged detail shown in the ring center displays the power density in
units of $I^2R_\square/\text{\textmu{}m}^2$ (see color scale on the right
bottom).
(b) Temperature profile along the blue line in (a) for various barrier
thicknesses.  Solid lines are calculated with Approx.\ I and dashed lines
with Approx.\ II.
(c) Temperature map of the ring structure with a barrier thickness of
$b=1\,$nm according to Approx.\ I.}
\end{figure}
Figure \ref{fig:v} (d) compares calculations for two different structures,
namely the four-probe configuration with source and drain contacts and two
voltage probes studied so far, and a two-probe configuration with no
voltage probes.  Clearly the temperature is reduced in the vicinity of the
current-free voltage probes.  However, their cooling effect is confined to
a small region whose extent is set by the relaxation parameter
$\lambda=\sqrt{bt\km/\kb}$ in Eq. \ref{equ:onedt}.  The design of cooling
fins would require to place them within a distance set by $\lambda$.  For
the model parameter used in our calculations  $\lambda\sim290\,$nm,
930\,nm, and 2.9\,\textmu{}m for $b=10\,$nm, 100\,nm, and 1\,\textmu{}m,
respectively.  The voltage probes in our test structure are within the
range of the central notch for $b=1\,$\textmu{}m, only.  Consequently, the
temperature is slightly reduced in the center in this case as can be seen
in Fig. \ref{fig:v} (d).  Note, however, that the temperature in the center
of the structure is peaking on top of the asymptotic temperature $T_a$ and
this temperature scales quadratically with $1/w$, the width of the source
and drain leads.  Thus, a stronger reduction can be obtained by making the
leads as wide as possible. 

To demonstrate the application of our numerical simulations for the
optimization of the sample layout with respect to control over the location
where gap formation sets in, we design a structure in which the current
path branches into a loop (see Fig.\ \ref{fig:vi}).  The final goal with
this specific structure is to produce two gaps, one in each loop arm.  If
bridged by molecules, such a sample could be used to test for the quantum
coherence of electron transport through molecules by measuring the
magnetoconductance.  Coherence would lead to the appearance of periodic
oscillation of the conductance in response to the magnetic flux penetrating
the area between the loop arms\cite{webb85}.  In the design of Fig.\
\ref{fig:vi} (a) we try to avoid unintentional current density enhancements
by rounding all concave corners.  A step in the width of the current path
is introduced only at two positions.  The inset in the ring center of Fig.\
\ref{fig:vi} (a) demonstrates that this leads to a considerable
current-density enhancement at singular points.  At the same time we would
like to have a large temperature gradient at this location.  To this end we
make the current path twice as narrow on the anode side of the ring
structure and thus the power density by a factor of four larger.  The
investigation of coherent transport requires a high aspect ratio of the
open area between the loop arms and the part of the ring area covered by
metal.  For this reason, the step in the ring width is placed close to the
cathode side of the structure.  Figure \ref{fig:vi} (c) shows the result of
Approx.\ I for a thin barrier ($b=1\,$nm).  The temperature changes
abruptly at the transition from the wide to the narrow part of the ring
thus making the concave corner at this position a preferred starting point
for void formation.  In Fig.\ \ref{fig:vi} (b) we display the temperature
profile along the path drawn as a blue line in Fig.\ \ref{fig:vi} (a) for
various barrier thicknesses.  The temperature gradient is large at point B
for a thin barrier as intended.  The temperature reaches to first
approximation almost constant values to the left of point A and between
point C and D for $b=1\,$nm on account of a short relaxation parameter
$\lambda\sim290\,$nm.  With increasing thickness $b$, however, Joule
heating leads to a more evenly distributed temperature due to the rise in
$\lambda\propto\sqrt{b}$.  As a result, the temperature gradient at point B
gets more and more shallow making the tendency to void formation at the
corner weaker.  Still, electromigration is boosted by the current-density
rise at this location which is thus distinguished from the rest of the
structure.  Nevertheless, the statistical nature of grain boundaries easily
marks different locations along the line from B to D where void formation
might start.  Hence, it is not clear from the outset that the corner close
to point B will win for larger $b$.  This again stresses that reliable void
formation is favored by good thermal coupling via a thin barrier to a
substrate of high thermal conductance.

\section{\label{sec:concl}Conclusion}
In this paper we have presented experimental results showing that nanogap
formation by electromigration depends crucially on the thickness of an
insulating barrier between the metallic structure and a thermally well
conducting substrate.  We have developed a numerical model which explains
these findings by considering the strength of thermal coupling of the
metallic structure to the substrate.  Only for thin barriers is cooling by
heat flow to the substrate efficient, while for thick barriers the
temperature is governed only by heat flow along the leads connecting the
microstructures to regions of lower current densities.  Thus, only thin
barriers allow the large current densities required for void formation on a
reasonable time scale without melting large parts of the structure by Joule
heating.  Therefore, good thermal coupling is an important prerequisite for
well controlled electromigration.  On the other hand, thin barriers are not
always an option in practice.  The native oxides on silicon wafers have
been reported to be as thin as $1\,$nm.  While such a thin barrier yields a
good thermal contact it bears the risk of undesirable electrical leakage
currents through the substrate.  For a reliable electrical insulation of
the structure a barrier of several 100\,nm thickness is mandatory.  In our
model calculation we considered cooling the structure by coupling it to a
cold substrate.  If electromigration has to be performed in vacuum for some
reason, this might be the only option.  In this case a thermally well
conducting insulator is required.  Another option would be to operate
electromigration in a controlled atmosphere.  We have, e.\,g., performed
electromigration experiments in the gas phase above the liquid level of a
Helium dewar.  The structure heated by a high current density leads to
convection of the cold gas which in turn gives rise to an efficient heat
release.  The yield of sufficiently small nanogaps considerably increased
in this case even on thick SiO$_2$ barriers.  Whether the resulting gaps
are clean and not contaminated with metallic nano-clusters is an open
question at this moment\cite{houck05,heersche06,zant06}.

Besides efficient heat release, a good layout design is needed to predefine
the position of void formation.  The latter starts preferably at positions
where high current densities meet large temperature gradients.  It is thus
advisable not to place the intended gap position in high symmetry points of
the layout where temperature gradients are expected to vanish but to mark
it with sharp corners which can be utilized for a considerable current
density enhancement. 

\begin{acknowledgments}
We acknowledge useful discussions with Regina Hoffmann, Dominik Stöffler,
and Michael Marz. 
\end{acknowledgments}

\appendix*

\section{\label{sec:app} Calculation of $G_b(\vecr)$}
We calculate $G_b(\vecr)$ by discrete cosine transform (DCT) on a cube of
$N\times{}N\times{}N_z$ cells (N=1024 turns out to be sufficient) with the
basic approach
\begin{widetext}\begin{equation}\label{equ:dct}
G_b(x,y,z)=\sum_{x,y=0}^{n-1}\sum_{z=0}^{N_z-1}\widehat{G}_b(X,Y,Z)
\cos(\pi x(X+1/2)/n)\cos(\pi y(Y+1/2)/n)\cos(\pi(z+1/2)(Z+1/2)/N_z),
\end{equation}\end{widetext}
where $n=N/2$ and the arguments of the cosine functions are chosen to
assure $G_b(\pm{}n,y,z)=G_b(x,\pm{}n,z)=G_b(x,y,N_z-1/2)=0$,
$G_b(1,y,z)=G_b(-1,y,z)$, $G_b(x,1,z)=G_b(x,-1,z)$, and
$G_b(x,y,0)=G_b(x,y,-1)$.  Together with the discrete version of the
Laplace equation,
\begin{equation}\label{equ:dlapl}
D^2G=\delta_{xyz},
\end{equation}
this describes a single source of unit strength at $(x=0,y=0,z=-1/2)$.
$D^2$ is the discrete version of the Laplace operator $\nabla^2$,
$\delta_{000}=1$, and $\delta_{xyz}=0$ for all other combinations of the
indices.  From Eq.\ (\ref{equ:dlapl}) one gets 
\begin{widetext}\begin{equation}\label{equ:dctnormal}
\widehat{G}_b(X,Y,Z)=\frac{\widehat{q}(X,Y,Z)}
{4(\cos(\pi(X+1/2)/n+\cos(\pi(Y+1/2)/n)+\cos(\pi(2Z+1)/N_z))-3}
\end{equation}\end{widetext}
where $\widehat{q}(X,Y,Z)$ is the DCT of $\delta_{xyz}$ defined with the
same arguments in the cosine functions as in Eq.\ (\ref{equ:dct}).  Thus
the calculation of $G_b(\vecr)$ involves two DCT operations and
normalization of the intermediate result according to Eq.\
(\ref{equ:dctnormal}).  The index $b$ enters into the calculation via the
definition of $N_z=b/h$, where $h$ is the discretization size in the
finite-difference method.  For Approx.\ I we need to know
$G(\vecr)=\lim_{b\to\infty}G_b(\vecr)$.  We approximate it by setting
$G(\vecr)\approx{}G_{512}$, which is sufficiently accurate due to the
asymptotic $1/r$ nature of the analytical solution.

\bibliography{kiessigetal}

\end{document}